\documentclass[sigconf]{acmart}

\usepackage{threeparttable}
\usepackage{glossaries}
\usepackage{xspace}
\setacronymstyle{long-short}
\newacronym{ott}{OTT}{Over-The-Top Media service}
\newacronym{vod}{VoD}{Video-on-Demand}
\newacronym{utc}{UTC}{Coordinated Universal Time}
\newacronym{vcpu}{vCPU}{virtual Central Processing Unit}
\newacronym[
longplural={User-Rating Matrices}
]
{urm}{URM}{User-Rating Matrix}

\newacronym{map}{MAP}{Mean Average Precision}
\newacronym{roc}{ROC}{Receiver Operating Characteristic curve}
\newacronym{rocauc}{ROC-AUC}{Receiver Operating Characteristic Area Under the Curve}
\newacronym{ndcg}{NDCG}{Normalized Discounted Cumulative Gain}
\newacronym{mrr}{MRR}{Mean Reciprocal Rank}
\newacronym{rs}{RS}{Recommender System}
\newacronym{ctr}{CTR}{Click Through Rate}

\AtBeginDocument{%
  \providecommand\BibTeX{{%
    \normalfont B\kern-0.5em{\scshape i\kern-0.25em b}\kern-0.8em\TeX}}}

\copyrightyear{2020}
\acmYear{2020}
\setcopyright{acmcopyright}\acmConference[CIKM '20]{Proceedings of the 29th ACM International Conference on Information and Knowledge Management}{October 19--23, 2020}{Virtual Event, Ireland}
\acmBooktitle{Proceedings of the 29th ACM International Conference on Information and Knowledge Management (CIKM '20), October 19--23, 2020, Virtual Event, Ireland}
\acmPrice{15.00}
\acmDOI{10.1145/3340531.3412774}
\acmISBN{978-1-4503-6859-9/20/10}

\acmSubmissionID{rt2300}

\begin{document}
\fancyhead{}

\newcommand{\dataset}{ContentWise Impressions\xspace}
\newcommand{\idest}{i.e.,\xspace}
\newcommand{\eg}{e.g.,\xspace}

\title{\dataset: An Industrial Dataset with Impressions Included}

\author{Fernando B. P\'erez Maurera}
\orcid{0000-0001-6578-7404}
\email{fernandobenjamin.perez@polimi.it}
\email{fernando.perez@contentwise.com}
\affiliation{%
  \institution{Politecnico di Milano, ContentWise}
  \city{Milan}
  \state{Italy}
}

\author{Maurizio Ferrari Dacrema}
\email{maurizio.ferrari@polimi.it}
\orcid{0000-0001-7103-2788}
\affiliation{%
  \institution{Politecnico di Milano}
  \streetaddress{Piazza Leonardo da Vinci, 32}
  \city{Milan}
  \state{Italy}
  \postcode{20133}
}

\author{Lorenzo Saule}
\email{lorenzo.saule@gmail.com}
\authornote{Intern at ContentWise and Ms.C. Student at Politecnico di Milano during the development of this work}
\orcid{0000-0002-1572-666X}
\affiliation{%
  \institution{Politecnico di Milano, ContentWise}
  \city{Milan}
  \state{Italy}
}

\author{Mario Scriminaci}
\email{mario.scriminaci@contentwise.com}
\affiliation{%
  \institution{ContentWise}
  \streetaddress{Via Simone Schiaffino, 11}
  \city{Milan}
  \state{Italy}
  \postcode{20158}
}

\author{Paolo Cremonesi}
\email{paolo.cremonesi@polimi.it}
\orcid{0000-0002-1253-8081} 
\affiliation{%
  \institution{Politecnico di Milano}
  \streetaddress{Piazza Leonardo da Vinci, 32}
  \city{Milan}
  \state{Italy}
  \postcode{20133}
}


\begin{abstract}
In this article, we introduce the \dataset dataset, a collection of implicit interactions and \emph{impressions} of movies and TV series from an Over-The-Top media service, which delivers its media contents over the Internet. The dataset is distinguished from other already available multimedia recommendation datasets by the availability of impressions, \idest the recommendations shown to the user, its size, and by being open-source. We describe the data collection process, the preprocessing applied, its characteristics, and statistics when compared to other commonly used datasets. We also highlight several possible use cases and research questions that can benefit from the availability of user impressions in an open-source dataset. Furthermore, we release software tools to load and split the data, as well as examples of how to use both user interactions and impressions in several common recommendation algorithms. 
\end{abstract}



\maketitle

\section{Introduction}
\label{sec:introduction}
Recommender Systems are, in this era of information, an ubiquitous technology that can be frequently found in the online services we use. The development of new algorithms and techniques has been fueled by the availability of public datasets to the community, collected by both researchers and industry.
 
The need to develop ever-better solutions is always present, driven by the evolution of business models and the availability of new data sources. An example of this is the RecSys Challenge\footnote{\url{https://recsys.acm.org/challenges/}} held every year since 2010. Each year an industry releases a dataset, challenging the participants on a problem that is relevant to their business model, \eg job recommendation, accommodation recommendation, playlist continuation. Other examples of these competitions are the KDD Cup\footnote{https://www.kdd.org/kdd2020/kdd-cup}, WSDM CUP\footnote{\url{http://www.wsdm-conference.org/2020/wsdm-cup-2020.php}}, and the several IJCAI\footnote{\url{https://www.ijcai20.org/competitions.html}} competitions. A recent emerging trend is to provide the \emph{impressions}, \idest what was recommended to the user alongside the user interactions. Recent articles, also from industry, propose algorithms that leverage impressions showing they can improve the recommendation quality \cite{wide-and-deep-learning-for-recommender-systems,click-through-rate-estimates-based-on-deep-learning,modeling-impression-discounting-in-large-scale-recommender-systems}.

Despite this growing research and industrial interest, as well as indications that impressions can be a useful information source, the research community is constrained by the lack of publicly available impression datasets. Most industrial datasets containing impressions have been released during challenges under a non-redistribute clause, or have been privately collected and mentioned in articles but never shared.

In order to address this limitation, in this work, we propose \emph{\dataset}, a new dataset of media recommendations containing impressions, that we release under a CC BY-NC-SA 4.0 license. We describe the data gathering process and provide statistics for the dataset comparing it to other commonly used datasets. We also provide further documentation and open-source tools to read the data, split it, and run several commonly used recommendation models.

The rest of this work is organized as follows. In Section~\ref{sec:state-of-the-art}, we provide an overview of other datasets with impressions. In Section~\ref{sec:data-description}, we describe \dataset. In Section~\ref{sec:dataset-building}, we explain the data gathering, preprocessing, and anonymization. In Section~\ref{sec:results}, we analyze the dataset and compare it with other datasets with impressions. In Section~\ref{sec:experiments}, we describe the experiments we performed and our observations from the evaluation procedures. Lastly, in Section~\ref{sec:conclusion}, we provide final remarks and provide future lines of work.

\section{Impressions datasets}
\label{sec:state-of-the-art}

Impression datasets have been used by several articles \cite{recsys-challenge-2016-job-recommendations,recsys-challenge-2017-offline-and-evaluation,recsys-challenge-2019-session-based-hotel-recommendations,modeling-impression-discounting-in-large-scale-recommender-systems,wide-and-deep-learning-for-recommender-systems,click-through-rate-estimates-based-on-deep-learning}. They can be classified into two categories: \emph{private} datasets, collected by the authors of the article but, to the best of our knowledge, not made accessible to the community, and \emph{non-redistributable} datasets, made accessible only to the participants of a challenge under a non-redistribute clause. In both cases, only a few researchers will have access to the dataset and will be able to use it. 
To the best of our knowledge, no open-source dataset with impressions exists.

\subsection{Private datasets}
Examples of private datasets are \emph{LinkedIn PYMK Impressions} and \emph{LinkedIn Skill Endorsement Impressions}. Both were used to model impressions discounting on large-scale \gls{rs}\cite{modeling-impression-discounting-in-large-scale-recommender-systems} and contain users registered on the LinkedIn\footnote{\url{https://www.linkedin.com/}} platform. 
More specifically, LinkedIn PYMK Impressions was used to recommend possible new user connections, and Linkedin Skill Endorsement was used to recommend skill endorsement of known users. Impressions in these datasets were present as a list of users, and a list of user-skill tuples, respectively \cite{modeling-impression-discounting-in-large-scale-recommender-systems}.

Another example of a private dataset is the mobile apps impressions used in \cite{wide-and-deep-learning-for-recommender-systems}. This dataset was gathered in order to develop a recommendation model for mobile applications on the Google Play Store in a low-latency scenario. In this dataset, impressions consist of mobile applications, application pages, historical statistics of the application, and more. 

\subsection{Non-redistributable datasets}
The impression datasets that have been made available to challenge participants under a non-redistribute clause in recent years are several. Examples are those provided during the RecSys Challenges: 2016 by Xing \cite{recsys-challenge-2016-job-recommendations, a-preliminary-study-on-a-recommender-system-for-the-job-recommendation-challenge}, 2017 by Xing \cite{recsys-challenge-2017-offline-and-evaluation}, and 2019 by Trivago\cite{recsys-challenge-2017-offline-and-evaluation}. Those datasets, however, were only accessible to participants of the challenge and have not been made available to the wider research community. 

Xing\footnote{\url{https://www.xing.com/}} is a social network for businesses, where users register to find jobs and recruiters register to find candidates. Users receive job recommendations. In Xing RecSys 2016, the impressions consist of the list of job recommendations provided to the user. 
In Xing RecSys 2017, the impressions were provided not as the recommendation list but rather as a boolean field to indicate if an item was shown to the user.

Trivago\footnote{\url{http://trivago.com/}} is a hotel search platform operating in several countries. In the Trivago RecSys 2019 dataset, users are provided with accommodation recommendations.

An older non-redistributable dataset is Tencent SearchAds Impressions \cite{modeling-impression-discounting-in-large-scale-recommender-systems, click-through-rate-estimates-based-on-deep-learning}, which was available during the KDD Cup 2012 Track 2\footnote{\url{https://www.kaggle.com/c/kddcup2012-track2}}. The dataset is a collection of interactions and impressions between users and the Tencent Search engine. The items in this dataset are represented by advertised results. The impressions comprise information about the user, the session, the query, the ads shown to the user, and their position on the screen.

\section{Data Description}
\label{sec:data-description}

In this section, we provide information about the data source and its content. The dataset is publicly available on Github\footnote{\url{https://github.com/ContentWise/contentwise-impressions}}.

\subsection{Source}

The data of \dataset comes from an \gls{ott}. This type of service offers media content to users via an Internet connection. In our case, the service offered content related to television and cinema. We collected the data for over four months between 2018 and 2019\footnote{Due to technical difficulties, there are certain days where no data is present}.

\begin{figure}[t]
  \centering
  \includegraphics[width=\linewidth]{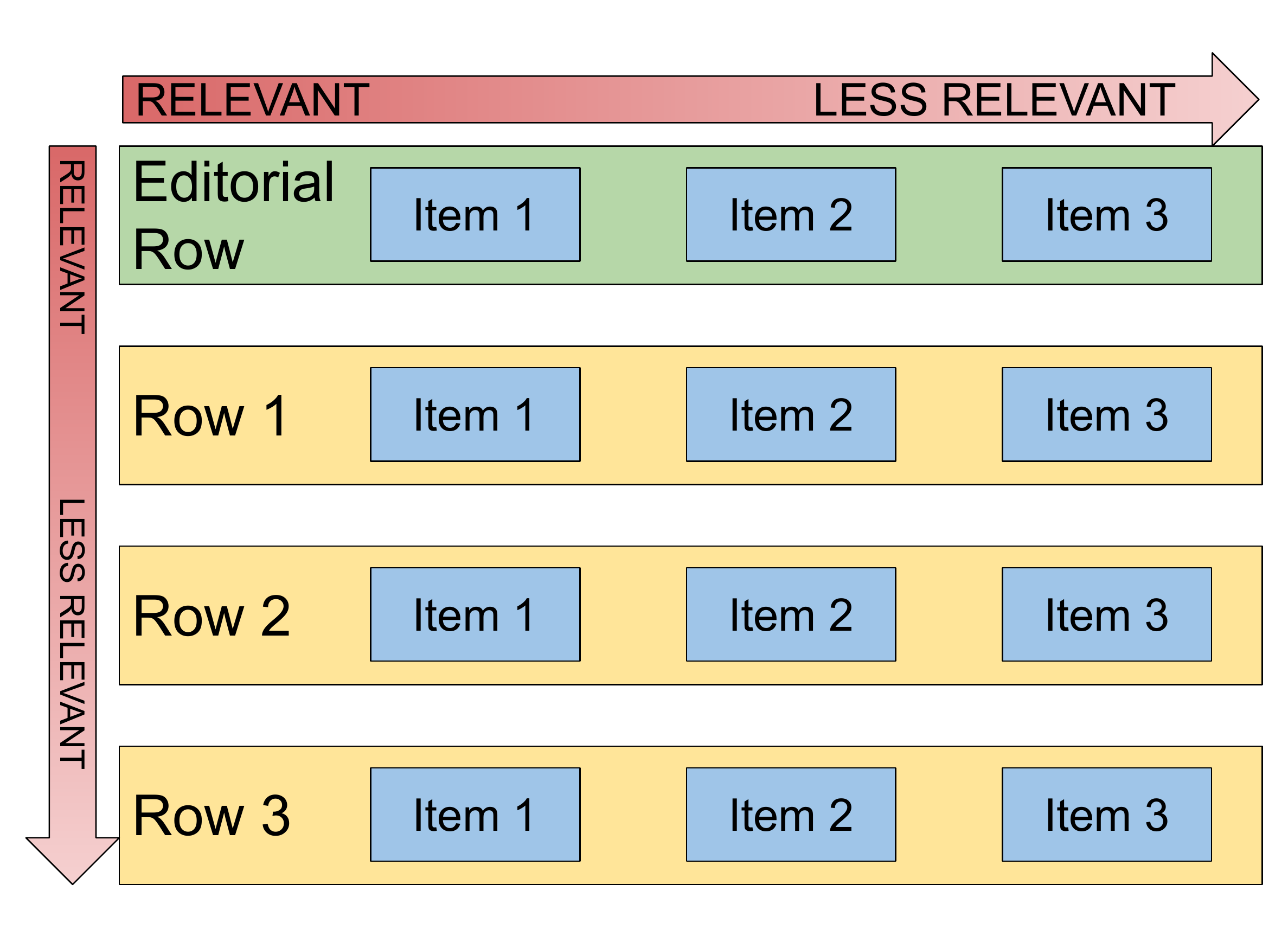}
  \caption{Example of the final user's screen layout. The \emph{editorial} recommendations are generated by the OTT, and are not included in this dataset. In this example, \dataset contains the impressions displayed in rows 1, 2, and 3.
  The number of rows and list lengths can vary. Most relevant rows were situated at the top of the screen. Most relevant items were situated to the left of the row.}
  \label{fig:screen-layout}
  \Description{Example of the service's layout.}
\end{figure}

In Figure~\ref{fig:screen-layout}, we report the screen layout shown to the users in which each row of the grid represents a recommendation list. The recommendations contained in a row are all generated by the same algorithm, but different rows can be generated by different algorithms, including non-personalized ones. Moreover, further rows could have been added by the service provider in between the rows we report in this dataset. Note that \dataset only contains the impressions that were provided by ContentWise while it does not contain those directly provided by the OTT.
 
\dataset is licensed under a CC BY-NC-SA 4.0 license\footnote{This license is available at \url{https://creativecommons.org/licenses/by-nc-sa/4.0}}. Moreover, we emphasize that it is explicitly forbidden to de-anonymize this dataset in order to link any of the identifiers to the original data source.

\subsection{Users}
Users represent registered accounts for the \gls{ott} service. Due to the nature of these types of services, several physical users (e.g., family members, friends) and devices can use the same account to interact with the service. Each user is represented by an anonymized numerical identifier.

\subsection{Items}
Items represent the multimedia content that the service provided to the users and are represented by an anonymized numerical identifier. As we mentioned before, all items are inside the media domain. More specifically, they refer to television and cinema products.

Items belong to four mutually exclusive categories: \emph{movies}, \emph{movies and clips in series}, \emph{TV movies or shows}, and  \emph{episodes of TV series}. These categories are encoded in the dataset with the values $0$, $1$, $2$, and $3$, respectively.
All items are associated with a \emph{series} identifier, which is used to group items belonging to the same series (i.e., TV series, movie series). Alongside this identifier, we also provide the \emph{episode number} and \emph{series length} for each item.

\subsection{Interactions}
The interactions represent the actions performed by users on items in the service and are associated with the timestamp\footnote{Specifically, the \gls{utc} UNIX timestamp, in milliseconds.} when it occurred. Interactions contain the identifier of the impressions, except in those cases where the recommendations came from a row added by the service provider. In Table~\ref{tab:interactions-columns}, we provide a description of the interaction data.
We categorized the interaction in four different types: \emph{views}, \emph{detail}, \emph{ratings}, and \emph{purchases}. These types are encoded in the dataset with the values $0$, $1$, $2$, and $3$, respectively.

\subsubsection{Views}
Views interactions indicate that the user watched an item of the service, and are represented with the interaction type \emph{zero}. We also provide, in the \emph{view factor} field, the point where the user stopped watching the item. The view factor is a real value that ranges from $0$ to $1$. If the user stopped watching near the end of the item, its view factor would be close to 1. On the other hand, if the user only viewed the beginning of the item, its view factor would be close to 0.

\subsubsection{Details}
Detail interactions indicate that the user accessed the item's detail page and are represented with the interaction type \emph{one}. 
    
\subsubsection{Purchases}
Some items need to be purchased before the user can watch them. Purchase interactions indicate that the user purchased an item of the catalog. We highlight that the catalog varied depending on the user's account subscription. Due to this, some users had to purchase items while others did not. The dataset does not contain any information about the user's account subscription.
  
\subsubsection{Rating}
Ratings are the only explicit feedback that the dataset contains, representing the rating value that a user gave to an item. Its values are in the range of 1-5 with a step of 0.5.

\subsection{Impressions}
\label{subsec:data-description:impressions}
The impressions refer to the recommended items that were presented to the user and are identified by their \emph{series}. Impressions consist of a numerical identifier, the list position on the screen, the length of the recommendation list, and an ordered list of recommended series identifiers, where the most relevant item is in the first position. We provide two types of impressions:

\subsubsection{Impressions with a direct link to interactions}
The user interacted with at least one item in the recommendation list. We identify these impressions with a numerical identifier. In Table~\ref{tab:impressions-direct-link-columns}, we describe the content of these impressions.

\subsubsection{Impressions without a direct link to interactions}
 
The user did not interact with any of the items in the recommendation list at the time the list was provided. Note that the user may have interacted with any of those items by other means, e.g., by successive recommendations, or search. We identify these impressions with the identifier of the user who received the recommendations. In Table~\ref{tab:impressions-non-direct-link-columns}, we describe the content of these impressions.

To summarize, \dataset is comprised of three different information layers. First, \emph{interactions} of users with items of the service, containing user-item pairs. Second, \emph{impressions with a direct link to interactions}, containing those recommendation lists that generated interactions. Third, \emph{impressions without a direct link to interactions}, containing those recommendation lists that did not generate interactions.

\subsection{Dataset format}
\label{subsec:dataset-format}

We provide the dataset as three different splits: \texttt{interactions}, \texttt{impressions-direct-link}, and \texttt{impressions-non-direct-link}. These are stored using the Apache's Parquet format\footnote{\url{https://parquet.apache.org/}}. This format is open source, data is stored in columns, and parsers can read and write data faster than classic $CSV$ parsers. There are several open-source tools for reading and writing Parquet files supporting several languages. We also include a human-readable $CSV$ version of the dataset to ensure long term availability of this resource.

In Table~\ref{tab:interactions-columns}, we provide the columns of the interactions and a description of them. All identifiers are anonymized, non-optional identifiers are always non-negative integers. Missing values are represented with $-1$. Similarly, in Table~\ref{tab:impressions-direct-link-columns} and Table~\ref{tab:impressions-non-direct-link-columns}, we describe the columns for both impression sets. All row positions are non-negative integers, recommendation lengths are positive integers, and the recommendation list contains at least one recommendation. 

\begin{table}
  \caption{Columns and their description for the interactions}
  \label{tab:interactions-columns}
  \begin{threeparttable}
    \begin{tabular}{lp{5cm}}
        \toprule
        \texttt{utc\_ts\_milliseconds}
            & UTC Unix timestamp of the interaction \\
        \texttt{user\_id}
            & Numerical identifier of users \\
        \texttt{item\_id}
            & Numerical identifier of items \\
        \texttt{series\_id}
            & Numerical identifier of series \\
        \texttt{recommendation\_id}
            & Optional numerical identifier of the impression. If the impression is not present, then its value is $-1$ \\
        \texttt{episode\_number}
            & Episode number of the item \\
        \texttt{series\_length}
            & Number of episodes of the series \\
        \texttt{item\_type}
            & Number to indicate the category of the item. Values range from 0 to 3 \\
        \texttt{interaction\_type}
            & Number to indicate the type of the interaction. Values range from 0 to 3 \\
        \texttt{explicit\_rating}
            & Rating value. If the interaction is not of type rating, then its value is $-1$ \\
        \texttt{vision\_factor}
            & Vision factor value. If the interaction is not of type view, then its value is $-1$ \\
        \bottomrule
    \end{tabular}
  \end{threeparttable}
\end{table}

\begin{table}
  \caption{Columns and their description for the impressions with a direct link to interactions.}
  \label{tab:impressions-direct-link-columns}
  \begin{threeparttable}
    \begin{tabular}{lp{4cm}}
        \toprule
        \texttt{recommendation\_id} \tnote{a}
            & Numerical identifier of the impression. \\
        \texttt{row\_position}
            & Position on screen of recommendation. \\
        \texttt{recommendation\_list\_length}
            & Number of recommended items \\
        \texttt{recommended\_series\_list} \tnote{b}
            & Ordered recommendation list of \texttt{series\_id}. \\
      \bottomrule
    \end{tabular}
    \begin{tablenotes}
        \item[a] This column is linked to the \texttt{recommendation\_id} column present in Table~\ref{tab:interactions-columns}
        
        \item[b] The series are linked to the \texttt{series\_id} column present in Table~\ref{tab:interactions-columns}
    \end{tablenotes}
  \end{threeparttable}
\end{table}

\begin{table}
  \caption{Columns and their description for the impressions without a direct link to interactions.}
  \label{tab:impressions-non-direct-link-columns}
  \begin{threeparttable}
    \begin{tabular}{lp{4cm}}
        \toprule
        \texttt{user\_id} \tnote{a}
            & Anonymized numerical identifier of the user that received the recommendation. \\
        \texttt{row\_position}
            & Position on screen of recommendation. \\
        \texttt{recommendation\_list\_length}
            & Number of recommended items \\
        \texttt{recommended\_series\_list} \tnote{b}
            & Ordered recommendation list of \texttt{series\_id}. \\
        \bottomrule
    \end{tabular}
    \begin{tablenotes}
        \item[a] This column is linked to the \texttt{user\_id} column present in Table~\ref{tab:interactions-columns}
        
        \item[b] The series are linked to the \texttt{series\_id} column present in Table~\ref{tab:interactions-columns}
    \end{tablenotes}
  \end{threeparttable}
\end{table}

\section{Dataset building}
\label{sec:dataset-building}

In this section, we describe the process to build the data from its source, passing through the preprocessing, and anonymization of it.

\subsection{Data acquisition}
\label{subsec:dataset-building:data-acquisition}

As mentioned in Section~\ref{sec:data-description}, {\dataset}' data comes from an \gls{ott} service, \idest on-demand media items, such as movies and TV series, which are streamed directly to the users via the Internet. We collected daily logs of interactions generated by the service and logs of recommendations made by our system. 

\subsection{Interactions preprocessing}
\label{subsec:dataset-building:interactions-preprocessing}

As a first preprocessing step, we removed users and interactions that had missing values due to technical issues. We also removed users that did not have any view interaction.

When a user started watching an item, a view interaction was generated with a starting vision factor. When the user finished watching the same item, another view interaction was generated, indicating the final vision factor. A small percentage of users had incorrect view factors (\eg always zero, no end view interaction) due to old software versions or technical issues. Users with invalid view factors have been removed.

Due to the significant size of the dataset, in order to make it suitable for research purposes, we built a subset of the original dataset via a uniform sampling of the users. The split we provide contains all interactions and impressions associated with the sampled users.

\subsection{Impressions preprocessing}
\label{sec:dataset-building:impressions-preprocessing}

In this section, we describe the preprocessing of the impressions related to the interactions previously selected.
Specifically, we grouped the impressions into the two disjoint sets described in Section \ref{subsec:data-description:impressions}: \emph{Impressions with a direct link to interactions}, when the user interacted with at least one of the recommended items, and \emph{Impressions without a direct link to interactions}, when the user did not interact with any recommendation.
Lastly, initial impressions logs did not contain the recommendation list length; we calculated and included these values.

\subsection{Data integrity}

The additional material we provide with this dataset also contains several tests to ensure the data integrity and its correspondence with the description provided here. For the impressions, we ensured all were of valid types and had a value in the correct ranges, rows had at least one item, reported row length was the same as the actual row length. Lastly, we ensured that row positions were always non-negative. The complete list of the integrity checks on \dataset is available in the online materials we provide.

\subsection{Anonymization}

A further preprocessing step involved the anonymization of all identifiers (\idest users, items, and series). Each of the original identifiers has been replaced with a random unique, and anonymous numerical identifier. This step is meant to make it impossible to reconstruct the user identity or to find their original accounts. Again, note that de-anonymizing the data is expressly forbidden. 

As mentioned in Section~\ref{sec:data-description}, the data has been collected over a period of four months between 2018 and 2019. The exact timestamps have been anonymized by applying a date and timezone shift. The day of the week has not been altered. After the date shift, the dataset contains timestamps from January, 7th 2019, to April, 15th 2019.
\section{Analysis and Discussion}
\label{sec:results}
In this section, we present an analysis of the \dataset and compare it with other datasets containing impressions.

\subsection{Analysis of the dataset}
\dataset contains $10,457,810$ interactions; $307,453$ impressions with direct links to interactions; and $23,342,617$ impressions without direct link to interactions. The dataset also contains $42,153$ users; $145,074$ items and $28,881$ series.

\begin{table}
  \caption{Number of interactions grouped by their type.}
  \label{tab:interaction-by-interaction-type}
  \begin{threeparttable}
    \begin{tabular}{p{3.5cm}rr}
        Interaction Type
            & \multicolumn{1}{c}{Count}
            & \multicolumn{1}{c}{Percentage}
            \\
        \toprule
        \emph{View}
            & $6,122,105$
            & $58.54\%$ \\
        \emph{Access}
            & $4,105,530$
            & $39.26\%$ \\
        \emph{Purchase}
            & $221,066$
            & $2.11\%$ \\
        \emph{Rating}
            & $9,109$
            & $0.09\%$ \\
        \bottomrule
        \textbf{Total}
            & $10,457,810$
            & $100\%$ \\
    \end{tabular}
  \end{threeparttable}
\end{table}

\begin{table}
  \caption{Number of interactions grouped by the item type.}
  \label{tab:interaction-by-item-type}
  \begin{threeparttable}
    \begin{tabular}{p{3.5cm}rr}
        Item Type
            & \multicolumn{1}{c}{Count}
            & \multicolumn{1}{c}{Percentage}
            \\
        \toprule
        \emph{Episodes of TV series}
            & $9,076,428$
            & $86.79\%$ \\
        \emph{Movies}
            & $987,518$
            & $9.44\%$ \\
        \emph{TV Movies and shows}
            & $162,574$
            & $1.56\%$ \\
        \emph{Movies and clips in series}
            & $231,290$
            & $2.21\%$ \\
        \bottomrule
        \textbf{Total}
            & $10,457,810$
            & $100\%$ \\
    \end{tabular}
  \end{threeparttable}
\end{table}

\begin{table}
  \caption{Number of items grouped by their type.}
  \label{tab:items-by-item-type}
  \begin{threeparttable}
    \begin{tabular}{p{3.9cm}rr}
        Item Type
            & \multicolumn{1}{c}{Count}
            & \multicolumn{1}{c}{Percentage}
            \\
        \toprule
        \emph{Episodes of TV series}
            & $123,831$
            & $85.36\%$ \\
        \emph{Movies}
            & $13,733$
            & $9.47\%$ \\
        \emph{TV Movies and shows}
            & $5,722$
            & $3.94\%$ \\
        \emph{Movies and clips in series}
            & $1,788$
            & $1.23\%$ \\
        \bottomrule
        \textbf{Total}
            & $145,074$
            & $100\%$ \\
    \end{tabular}
  \end{threeparttable}
\end{table}

In Table~\ref{tab:interaction-by-interaction-type}, we highlight the distribution of the interactions when grouped by interaction type, where $97.8\%$ of the dataset is comprised of \emph{view} and \emph{access} interactions. Similarly, in Table~\ref{tab:interaction-by-item-type}, we present the distribution of interactions by item type, where $96.23\%$ of the interactions correspond to \emph{episodes of TV series} and \emph{movies}. Lastly, in Table~\ref{tab:items-by-item-type}, we show the distribution of item types, where the same \emph{episodes of TV series} and \emph{movies} item types represent $94.83\%$ of the total items.

We observed that users, items, and series, present long-tail distributions. For users, $27.96\%$ most popular users are associated with $80\%$ of the interactions. For items, $12.06\%$ most popular items correspond with $80\%$ of the interactions. For series, $4.05\%$ most popular series appear in $80\%$ of the interactions. 

The average number of interactions per user is $248$ ($22$ if counting direct interactions from impressions), where the maximum and the minimum number of interactions made by a single user are $13,517$ and $2$ ($2,886$ and $1$ if counting direct interactions from impressions), respectively.

For items, the average number of interactions received per item is $72$ ($25$ if counting interactions from impressions), where the maximum and the minimum number of interactions received by a single item are $23,939$ and $1$ ($6,260$ and $1$ if counting interactions from impressions), respectively.

For impressions with direct links to interactions, the average number of interactions received per impression is $2$, where the maximum and the minimum number of interactions received by a single item are $213$ and $1$, respectively. 

\begin{figure}[t]
  \centering
  \includegraphics[width=\linewidth]{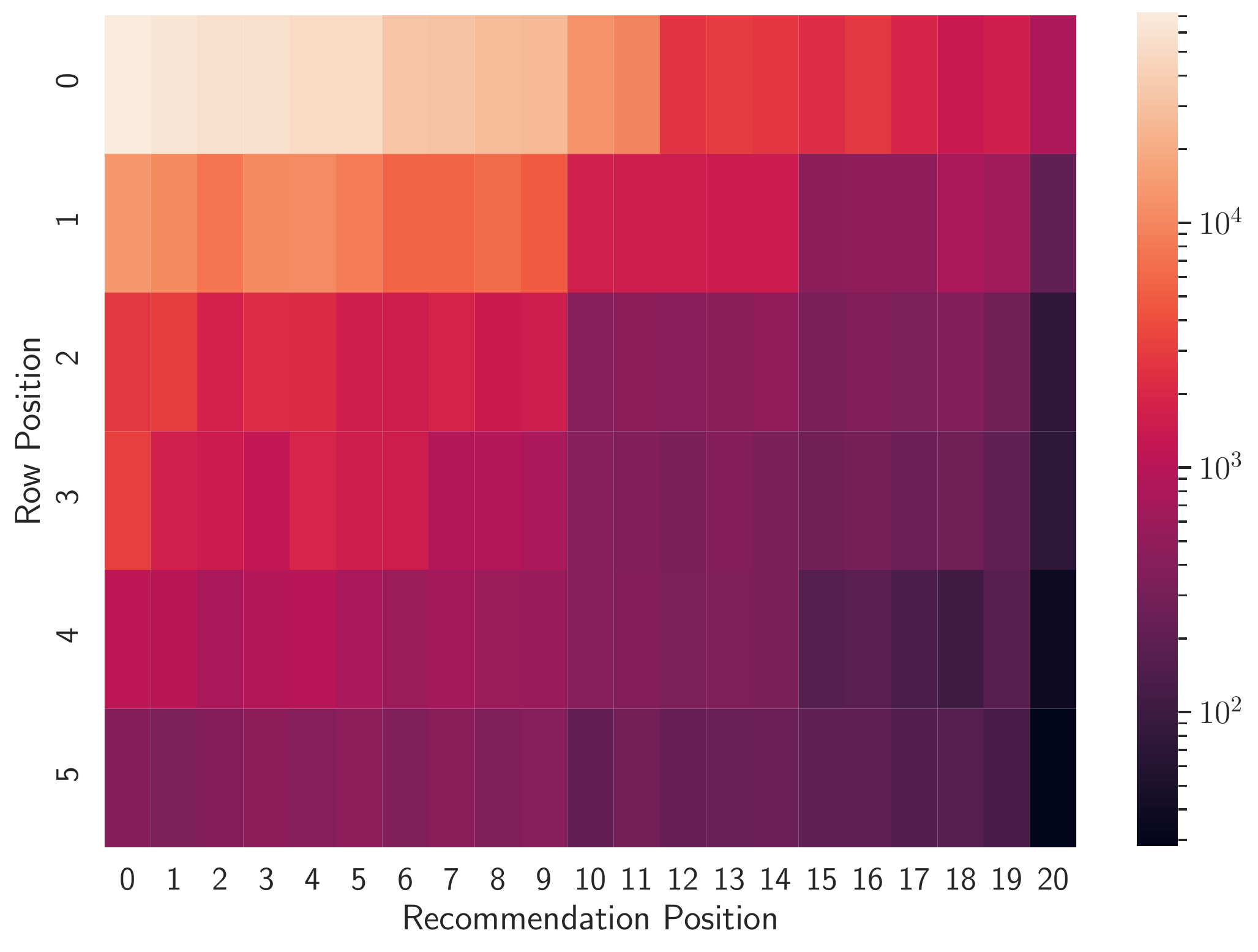}
  \caption{Heatmap of the number of interactions per position on the screen. Most interacted items are located in the first rows and on the first positions of the list. Values are log-scaled.}
  \label{fig:impressions-heatmap}
  \Description{Heatmap of the number of interactions per row position and position in recommendation list}
\end{figure}

In Figure~\ref{fig:impressions-heatmap}, we show a heatmap that indicates the most interacted positions of the recommendation lists based on the row position on the screen. Specifically, we see that most interactions happen between the first three row positions, and the first ten item positions.

\subsection{Comparison with other datasets}
As previously mentioned in Section~\ref{sec:state-of-the-art}, currently, no impressions datasets are publicly available to the community. As such, we gathered and reported their statistics using the ones described on works that used those datasets.

To the best of our knowledge, \dataset is the first dataset with impressions to be open-sourced. In previous years, other articles have used private datasets\cite{modeling-impression-discounting-in-large-scale-recommender-systems,wide-and-deep-learning-for-recommender-systems}, which were not released to the community. Others were disclosed under non-redistribution clauses on challenges\cite{recsys-challenge-2016-job-recommendations, recsys-challenge-2017-offline-and-evaluation, recsys-challenge-2019-session-based-hotel-recommendations, click-through-rate-estimates-based-on-deep-learning}, where only a few researchers have access to them. 
Furthermore, \dataset provides both impressions present in the interactions and without any associated interaction. Both LinkedIn PYMK Impressions and LinkedIn Skill Endorsement \cite{modeling-impression-discounting-in-large-scale-recommender-systems} also present both impressions. On the other hand, other datasets \cite{recsys-challenge-2016-job-recommendations,recsys-challenge-2019-session-based-hotel-recommendations} only provided impressions present in the interactions.

Another advantage of \dataset is that it is subsampled in a way to be easily usable for research purposes without requiring significant computation resources. While researchers can indeed preprocess and subsample bigger datasets, if needed, this may result in different articles relying on different subsampling, making it more difficult to compare research results and contributing to the reproducibility crisis in our field \cite{are-we-really-making-much-progress-a-worrying-analysis-of-recent-neural-recommendation-approaches,dacrema2019troubling}.
For instance, Xing RecSys 2017\cite{recsys-challenge-2016-job-recommendations,recsys-challenge-2017-offline-and-evaluation} contained around $1.5M$ users, $1.3M$ items, $322M$ interactions, and $314M$ impressions associated with these interactions. LinkedIn PYMK Impressions, LinkedIn Skill Endorsement Impressions, Tencent SearchAds Impressions \cite{modeling-impression-discounting-in-large-scale-recommender-systems} had $1.08$, $0.19$, and $0.15$ billion impressions. For comparison, commonly used research datasets have a number of users and items in the range of tens of thousands and up to a few millions of interactions \cite{dacrema2019troubling}.

In Table~\ref{tab:dataset-comparison}, we provide a comparison of the density and Gini indexes of the datasets we could obtain\footnote{We could not acquire any of the private datasets and therefore could not compute those additional statistics.}. The Gini Index is computed on the number of interactions associated with each user or on those associated with each item. As a reference, we also provide these values for the more-commonly used MovieLens 20M dataset\cite{movielens}. From Table~\ref{tab:dataset-comparison}, we can see that \dataset is significantly denser than other impression datasets, while sparser than Movielens. In terms of the Gini Index, higher values indicate the dataset is more biased towards popular items or users with long profiles. \dataset exhibits a significantly lower popularity bias on both the items and the users, indicating that the data is more balanced.

\begin{table}[ht]
  \caption{Comparison of \dataset with others datasets based on their density, Gini indexes on item popularity and users.}
  \label{tab:dataset-comparison}
  \begin{tabular}{lrcc}
    \toprule
    Dataset
        & \multicolumn{1}{c}{Density}
        & \multicolumn{1}{c}{Gini Items}
        & \multicolumn{1}{c}{Gini Users}
        \\
    \midrule
    \dataset
        & $7.4 \cdot 10^{-4}$
        & $0.3345$
        & $0.3316$ \\
    Xing Recsys 2016
        & $6.4 \cdot 10^{-6}$
        & $0.6890$
        & $0.7652$ \\ 
    Xing Recsys 2017
        & $6.6 \cdot 10^{-6}$
        & $0.6530$
        & $0.9241$ \\ 
    Movielens 20M
        & $5.3 \cdot 10^{-3}$
        & $0.5807$	
        & $0.9048$ \\
  \bottomrule
\end{tabular}
\end{table}

\section{Experiments}
\label{sec:experiments}
The purpose of the experiments is both to report baseline results for the \dataset as well as provide examples in the online materials that researchers can use and refer to. In this section, we describe the experiments we performed on the dataset. 

We provide open-source materials written in Python to download the dataset, install the environment, read the data, parse it, and use several common recommendation models. The source code is available on Github\footnote{\url{https://github.com/ContentWise/contentwise-impressions}}. The source code relies on common open-sourced scientific libraries. We ran the experiments on a single Linux Amazon EC2 {r4.4xlarge} instance. At the time of writing, this type of instance provides 16 vCPU and 128 GiB of RAM. 

\subsection{Recommendation task}
We evaluated the models under a traditional \emph{top-k} recommendation task with only collaborative information, \idest user-item interactions and impressions.
We rely on the publicly available evaluation framework\footnote{\url{https://github.com/MaurizioFD/RecSys2019_DeepLearning_Evaluation}} developed by Ferrari Dacrema et al. \cite{are-we-really-making-much-progress-a-worrying-analysis-of-recent-neural-recommendation-approaches,dacrema2019troubling}. We added a few changes to the framework in order to support the parallel evaluation of users and utility methods to extract, transform, and load \dataset. We also included consistency checks of the dataset using unit tests.

The data is split via random holdout of the interactions in training ($70\%$), validation ($10\%$) and test ($20\%$). All interactions are considered as implicit with a value of 1.

\subsection{Baseline algorithms}
\label{subsec:related-works:recommendation-techniques}
We report the recommendation quality of several simple algorithms. As non-personalized baselines, we report a \emph{Top Popular} recommender, which recommends the items having the highest number of interactions. As the personalized recommenders, we report \emph{ItemKNN}, a simple neighborhood-based recommender using various similarities measures:  cosine\cite{data-mining-methods-for-recommender-systems}, dice\cite{dice-similarity-1, dice-similarity-2}, jaccard\cite{data-mining-methods-for-recommender-systems}, asymmetric\cite{efficient-top-n-recommendation-for-very-large-scale-binary-rated-datasets}, and tversky\cite{tversky-features-of-similarity}. We also report a graph-based algorithm $RP^{3}_{\beta}$ proposed in \cite{updatable-accurate-diverse-and-scalable-recommendations-for-interactive-applications}. 
For latent-factor methods, we report \emph{PureSVD} \cite{performace-of-recommender-algorithms-on-top-n-recommendation-tasks} and \emph{MF BPR} \cite{bpr-bayesian-personalized-ranking-from-implicit-feedback}.

In order to provide a simple example of how to use the impressions during the training phase of a model, we adapted the MF BPR algorithm. Traditional BPR requires to sample for each user, a positive interaction (\idest an item the user interacted with), and a negative interaction (\idest an item the user did not interact with). Specifically, we did not alter the positive sampling, but we experimented with three different strategies for the negative sampling: \emph{uniform-at-random}, sampling uniformly among the items the user did not interact with; \emph{uniform-inside-impressions}, sampling uniformly among the user impressions; and \emph{uniform-outside-impressions}, sampling uniformly among the items not in the impressions. Items the user interacted with are never sampled as negatives.

\subsubsection{Hyperparameter tuning}
We tuned the hyperparameters of each recommendation algorithm, optimizing the recommendation quality on the validation data. We applied Bayesian Optimization\cite{a-tutorial-on-bayesian-optimization-of-expensive-cost-functions,antenucci2018artist} and set the hyperparameter ranges and distributions according to those used in \cite{are-we-really-making-much-progress-a-worrying-analysis-of-recent-neural-recommendation-approaches,dacrema2019troubling}. When the Bayesian search ended, we trained the algorithm using the best hyperparameter found on the union of train and validation data and report the results obtained on the test data.

\subsubsection{Evaluation}

We measured the performance of the recommendation techniques using both accuracy and beyond-accuracy metrics at recommendation list lengths 20. We report results of Precision, \gls{map}, \gls{ndcg}, and Item Coverage (Item Cov), which represents the quota of items that were recommended at least once\footnote{We exported results with more metrics in the repository.}.

\subsection{Experiments result}

In Table~\ref{tab:all-metrics}, we report the results of the evaluation. We can observe that the best performing algorithm is ItemKNN, in particular with the tversky similarity. Other algorithms, like $RP^{3}_{\beta}$ and PureSVD, have a lower recommendation quality. The recommendation quality of the simple MF BPR baseline is relatively low, achieving a similar recommendation quality as the Top Popular baseline. This suggests the need for further studies to develop a more suitable algorithmic solution.

When comparing the MF BPR negative items sampling strategies, we can observe that the recommendation quality overall does not change dramatically but shows a tendency to decrease in both cases when impressions are used. The recommendation quality decreases the most when negative items are sampled within the impressions. This behavior is expected for two reasons. First, the impressions are the recommendations that were provided to a user by another recommendation model. Therefore, they are unlikely to contain strongly negative items. Sampling negative items among impressions will result in considering as negatives those items that are close to the interests of the user, therefore steering the algorithm in the wrong direction. Second, sampling only outside of the impressions is, too, a limited strategy, as erroneous recommendations will not be sampled as negatives and will prevent the algorithm to further refine its quality. Both these results indicate that a more articulate sampling strategy can be developed, potentially merging the strengths of the two, while minimizing their weaknesses.

As another interesting observation, we can see that the Item Coverage of the MF BPR algorithm is much better than the Top Popular one, indicating that despite its similar recommendation quality, the MF BPR allows for a far greater exploration of the catalog. In this case, we can see a significant difference between negative sampling strategies. Sampling negatives within the impressions results in a markedly low item coverage, whereas sampling outside the impressions allows the model to improve the item coverage over the plain uniform negative sampling.

\begin{table}[ht]
    \caption{Evaluation of different metrics on recommendation lists of length 20. Best results highlighted in bold.}
    \label{tab:all-metrics}
    \begin{tabular}{lcccc}
        \toprule
        {} 
        {} 
            &    PREC
            &     MAP
            &    NDCG
            & \begin{tabular}{@{}c@{}}Cov. \\ Item\end{tabular} \\
        \midrule
        TopPop                
            &  0.0225 
            &  0.0387 
            &  0.0619 
            & 0.0006 \\
        \midrule
        ItemKNN CF cosine     
            &  0.2562 
            &  0.3972 
            &  0.4907 
            & 0.3431 \\
        ItemKNN CF dice       
            &  0.2565 
            &  0.3952 
            &  0.4878 
            & 0.3887 \\
        ItemKNN CF jaccard    
            &  0.2574 
            &  0.3979 
            &  0.4910 
            & 0.4203 \\ 
        ItemKNN CF asymmetric 
            &  0.2549 
            &  0.3949 
            &  0.4896 
            & 0.3225 \\
        ItemKNN CF tversky    
            &  \textbf{0.2587} 
            &  \textbf{0.4010} 
            &  \textbf{0.4935} 
            & 0.3791 \\
        RP3beta               
            &  0.1687 
            &  0.2641  
            &  0.3664 
            & 0.3502 \\
        \midrule
        MF BPR                
            &  0.0314 
            &  0.0531 
            &  0.0900
            & 0.1012 \\
        MF BPR inside         
            &  0.0205
            &  0.0323
            &  0.0550 
            & 0.0006 \\
        MF BPR outside        
            &  0.0195
            &  0.0395
            &  0.0619
            & 0.1202 \\
        PureSVD               
            &  0.1730
            &  0.2416
            &  0.3369
            &  0.0897 \\
        \bottomrule
\end{tabular}
\end{table}

\section{Conclusion and Future Works}
\label{sec:conclusion}

In this work, we presented \dataset, a novel dataset with impressions, gathered from an industrial service provider which, to the best of our knowledge, is the first one to be publicly available to the research community. The dataset is licensed under a CC BY-NC-SA 4.0 license, allowing its wide usage for both academic and industry research.

We described the contents of the dataset, from its users, items, interactions, impressions, and format. We also documented how we built it, going from its source, preprocessing, and finally, its anonymization. We analyzed the dataset, compared it against other datasets, and presented the results of our experiments. In these, we observed how the use of impressions affects the performance of some state-of-the-art recommendation techniques. We open-sourced all the tools and documentation that we used so others can reproduce our observations. Moreover, inside these tools, we provided instructions to download, load, and use the dataset. 

\dataset can enable the community to further study how to embed the impression information in algorithmic solutions for recommendations. Possible research directions are, for example, refining the user model according to how many times they did not interact with a recommended item, when to stop to recommend an item to a user, reranking strategies to compensate known errors that the recommendation model generating the impressions has been found to make.
Another possibility is to post-process the recommendations in order to mitigate biases that the impressions may exhibit. Lastly, if met with interest from the community, updated and bigger versions of this dataset can be released in the future.

\bibliographystyle{ACM-Reference-Format}
\bibliography{main}


\begin{thebibliography}{20}


\ifx \showCODEN    \undefined \def \showCODEN     #1{\unskip}     \fi
\ifx \showDOI      \undefined \def \showDOI       #1{#1}\fi
\ifx \showISBNx    \undefined \def \showISBNx     #1{\unskip}     \fi
\ifx \showISBNxiii \undefined \def \showISBNxiii  #1{\unskip}     \fi
\ifx \showISSN     \undefined \def \showISSN      #1{\unskip}     \fi
\ifx \showLCCN     \undefined \def \showLCCN      #1{\unskip}     \fi
\ifx \shownote     \undefined \def \shownote      #1{#1}          \fi
\ifx \showarticletitle \undefined \def \showarticletitle #1{#1}   \fi
\ifx \showURL      \undefined \def \showURL       {\relax}        \fi
\providecommand\bibfield[2]{#2}
\providecommand\bibinfo[2]{#2}
\providecommand\natexlab[1]{#1}
\providecommand\showeprint[2][]{arXiv:#2}

\bibitem[\protect\citeauthoryear{Abel, Bencz\'{u}r, Kohlsdorf, Larson, and
  P\'{a}lovics}{Abel et~al\mbox{.}}{2016}]%
        {recsys-challenge-2016-job-recommendations}
\bibfield{author}{\bibinfo{person}{Fabian Abel}, \bibinfo{person}{Andr\'{a}s
  Bencz\'{u}r}, \bibinfo{person}{Daniel Kohlsdorf}, \bibinfo{person}{Martha
  Larson}, {and} \bibinfo{person}{R\'{o}bert P\'{a}lovics}.}
  \bibinfo{year}{2016}\natexlab{}.
\newblock \showarticletitle{RecSys Challenge 2016: Job Recommendations}. In
  \bibinfo{booktitle}{\emph{Proceedings of the 10th ACM Conference on
  Recommender Systems}} (Boston, Massachusetts, USA)
  \emph{(\bibinfo{series}{RecSys '16})}. \bibinfo{publisher}{Association for
  Computing Machinery}, \bibinfo{address}{New York, NY, USA},
  \bibinfo{pages}{425--426}.
\newblock
\showISBNx{9781450340359}
\urldef\tempurl%
\url{https://doi.org/10.1145/2959100.2959207}
\showDOI{\tempurl}


\bibitem[\protect\citeauthoryear{Abel, Deldjoo, Elahi, and Kohlsdorf}{Abel
  et~al\mbox{.}}{2017}]%
        {recsys-challenge-2017-offline-and-evaluation}
\bibfield{author}{\bibinfo{person}{Fabian Abel}, \bibinfo{person}{Yashar
  Deldjoo}, \bibinfo{person}{Mehdi Elahi}, {and} \bibinfo{person}{Daniel
  Kohlsdorf}.} \bibinfo{year}{2017}\natexlab{}.
\newblock \showarticletitle{RecSys Challenge 2017: Offline and Online
  Evaluation}. In \bibinfo{booktitle}{\emph{Proceedings of the Eleventh ACM
  Conference on Recommender Systems}} (Como, Italy)
  \emph{(\bibinfo{series}{RecSys '17})}. \bibinfo{publisher}{Association for
  Computing Machinery}, \bibinfo{address}{New York, NY, USA},
  \bibinfo{pages}{372--373}.
\newblock
\showISBNx{9781450346528}
\urldef\tempurl%
\url{https://doi.org/10.1145/3109859.3109954}
\showDOI{\tempurl}


\bibitem[\protect\citeauthoryear{Aiolli}{Aiolli}{2013}]%
  {efficient-top-n-recommendation-for-very-large-scale-binary-rated-datasets}
\bibfield{author}{\bibinfo{person}{Fabio Aiolli}.}
  \bibinfo{year}{2013}\natexlab{}.
\newblock \showarticletitle{Efficient Top-n Recommendation for Very Large Scale
  Binary Rated Datasets}. In \bibinfo{booktitle}{\emph{Proceedings of the 7th
  ACM Conference on Recommender Systems}} (Hong Kong, China)
  \emph{(\bibinfo{series}{RecSys '13})}. \bibinfo{publisher}{Association for
  Computing Machinery}, \bibinfo{address}{New York, NY, USA},
  \bibinfo{pages}{273--280}.
\newblock
\showISBNx{9781450324090}
\urldef\tempurl%
\url{https://doi.org/10.1145/2507157.2507189}
\showDOI{\tempurl}


\bibitem[\protect\citeauthoryear{Amatriain and Pujol}{Amatriain and
  Pujol}{2015}]%
        {data-mining-methods-for-recommender-systems}
\bibfield{author}{\bibinfo{person}{Xavier Amatriain} {and}
  \bibinfo{person}{Josep~M. Pujol}.} \bibinfo{year}{2015}\natexlab{}.
\newblock \bibinfo{booktitle}{\emph{Data Mining Methods for Recommender
  Systems}}.
\newblock \bibinfo{publisher}{Springer US}, \bibinfo{address}{Boston, MA},
  \bibinfo{pages}{227--262}.
\newblock
\showISBNx{978-1-4899-7637-6}
\urldef\tempurl%
\url{https://doi.org/10.1007/978-1-4899-7637-6_7}
\showDOI{\tempurl}


\bibitem[\protect\citeauthoryear{Antenucci, Boglio, Chioso, Dervishaj, Kang,
  Scarlatti, and Dacrema}{Antenucci et~al\mbox{.}}{2018}]%
        {antenucci2018artist}
\bibfield{author}{\bibinfo{person}{Sebastiano Antenucci},
  \bibinfo{person}{Simone Boglio}, \bibinfo{person}{Emanuele Chioso},
  \bibinfo{person}{Ervin Dervishaj}, \bibinfo{person}{Shuwen Kang},
  \bibinfo{person}{Tommaso Scarlatti}, {and} \bibinfo{person}{Maurizio~Ferrari
  Dacrema}.} \bibinfo{year}{2018}\natexlab{}.
\newblock \showarticletitle{Artist-driven Layering and User's Behaviour Impact
  on Recommendations in a Playlist Continuation Scenario}. In
  \bibinfo{booktitle}{\emph{Recommender Systems Challenge Workshop at the 12th
  ACM Conference on Recommender Systems}}. \bibinfo{pages}{4:1--4:6}.
\newblock


\bibitem[\protect\citeauthoryear{Brochu, Cora, and de~Freitas}{Brochu
  et~al\mbox{.}}{2010}]%
        {a-tutorial-on-bayesian-optimization-of-expensive-cost-functions}
\bibfield{author}{\bibinfo{person}{Eric Brochu}, \bibinfo{person}{Vlad~M.
  Cora}, {and} \bibinfo{person}{Nando de Freitas}.}
  \bibinfo{year}{2010}\natexlab{}.
\newblock \showarticletitle{A Tutorial on Bayesian Optimization of Expensive
  Cost Functions, with Application to Active User Modeling and Hierarchical
  Reinforcement Learning}.
\newblock \bibinfo{journal}{\emph{CoRR}}  \bibinfo{volume}{abs/1012.2599}
  (\bibinfo{year}{2010}).
\newblock
\showeprint[arxiv]{1012.2599}
\urldef\tempurl%
\url{http://arxiv.org/abs/1012.2599}
\showURL{%
\tempurl}


\bibitem[\protect\citeauthoryear{Cheng, Koc, Harmsen, Shaked, Chandra, Aradhye,
  Anderson, Corrado, Chai, Ispir, Anil, Haque, Hong, Jain, Liu, and Shah}{Cheng
  et~al\mbox{.}}{2016}]%
        {wide-and-deep-learning-for-recommender-systems}
\bibfield{author}{\bibinfo{person}{Heng-Tze Cheng}, \bibinfo{person}{Levent
  Koc}, \bibinfo{person}{Jeremiah Harmsen}, \bibinfo{person}{Tal Shaked},
  \bibinfo{person}{Tushar Chandra}, \bibinfo{person}{Hrishi Aradhye},
  \bibinfo{person}{Glen Anderson}, \bibinfo{person}{Greg Corrado},
  \bibinfo{person}{Wei Chai}, \bibinfo{person}{Mustafa Ispir},
  \bibinfo{person}{Rohan Anil}, \bibinfo{person}{Zakaria Haque},
  \bibinfo{person}{Lichan Hong}, \bibinfo{person}{Vihan Jain},
  \bibinfo{person}{Xiaobing Liu}, {and} \bibinfo{person}{Hemal Shah}.}
  \bibinfo{year}{2016}\natexlab{}.
\newblock \showarticletitle{Wide \& Deep Learning for Recommender Systems}. In
  \bibinfo{booktitle}{\emph{Proceedings of the 1st Workshop on Deep Learning
  for Recommender Systems}} (Boston, MA, USA) \emph{(\bibinfo{series}{DLRS
  2016})}. \bibinfo{publisher}{Association for Computing Machinery},
  \bibinfo{address}{New York, NY, USA}, \bibinfo{pages}{7--10}.
\newblock
\showISBNx{9781450347952}
\urldef\tempurl%
\url{https://doi.org/10.1145/2988450.2988454}
\showDOI{\tempurl}


\bibitem[\protect\citeauthoryear{Cremonesi, Koren, and Turrin}{Cremonesi
  et~al\mbox{.}}{2010}]%
        {performace-of-recommender-algorithms-on-top-n-recommendation-tasks}
\bibfield{author}{\bibinfo{person}{Paolo Cremonesi}, \bibinfo{person}{Yehuda
  Koren}, {and} \bibinfo{person}{Roberto Turrin}.}
  \bibinfo{year}{2010}\natexlab{}.
\newblock \showarticletitle{Performance of Recommender Algorithms on Top-n
  Recommendation Tasks}. In \bibinfo{booktitle}{\emph{Proceedings of the Fourth
  ACM Conference on Recommender Systems}} (Barcelona, Spain)
  \emph{(\bibinfo{series}{RecSys '10})}. \bibinfo{publisher}{Association for
  Computing Machinery}, \bibinfo{address}{New York, NY, USA},
  \bibinfo{pages}{39--46}.
\newblock
\showISBNx{9781605589060}
\urldef\tempurl%
\url{https://doi.org/10.1145/1864708.1864721}
\showDOI{\tempurl}


\bibitem[\protect\citeauthoryear{Dacrema, Cremonesi, and Jannach}{Dacrema
  et~al\mbox{.}}{2019}]%
  {are-we-really-making-much-progress-a-worrying-analysis-of-recent-neural-recommendation-approaches}
\bibfield{author}{\bibinfo{person}{Maurizio~Ferrari Dacrema},
  \bibinfo{person}{Paolo Cremonesi}, {and} \bibinfo{person}{Dietmar Jannach}.}
  \bibinfo{year}{2019}\natexlab{}.
\newblock \showarticletitle{Are We Really Making Much Progress? A Worrying
  Analysis of Recent Neural Recommendation Approaches}. In
  \bibinfo{booktitle}{\emph{Proceedings of the 13th ACM Conference on
  Recommender Systems}} (Copenhagen, Denmark) \emph{(\bibinfo{series}{RecSys
  '19})}. \bibinfo{publisher}{Association for Computing Machinery},
  \bibinfo{address}{New York, NY, USA}, \bibinfo{pages}{101--109}.
\newblock
\showISBNx{9781450362436}
\urldef\tempurl%
\url{https://doi.org/10.1145/3298689.3347058}
\showDOI{\tempurl}


\bibitem[\protect\citeauthoryear{Dice}{Dice}{1945}]%
        {dice-similarity-2}
\bibfield{author}{\bibinfo{person}{Lee~R. Dice}.}
  \bibinfo{year}{1945}\natexlab{}.
\newblock \showarticletitle{Measures of the Amount of Ecologic Association
  Between Species}.
\newblock \bibinfo{journal}{\emph{Ecology}} \bibinfo{volume}{26},
  \bibinfo{number}{3} (\bibinfo{year}{1945}), \bibinfo{pages}{297--302}.
\newblock
\urldef\tempurl%
\url{https://doi.org/10.2307/1932409}
\showDOI{\tempurl}
\showeprint{https://esajournals.onlinelibrary.wiley.com/doi/pdf/10.2307/1932409}


\bibitem[\protect\citeauthoryear{Ferrari~Dacrema, Boglio, Cremonesi, and
  Jannach}{Ferrari~Dacrema et~al\mbox{.}}{2019}]%
        {dacrema2019troubling}
\bibfield{author}{\bibinfo{person}{Maurizio Ferrari~Dacrema},
  \bibinfo{person}{Simone Boglio}, \bibinfo{person}{Paolo Cremonesi}, {and}
  \bibinfo{person}{Dietmar Jannach}.} \bibinfo{year}{2019}\natexlab{}.
\newblock \showarticletitle{A Troubling Analysis of Reproducibility and
  Progress in Recommender Systems Research}.
\newblock \bibinfo{journal}{\emph{arXiv:1911.07698}} (\bibinfo{year}{2019}).
\newblock


\bibitem[\protect\citeauthoryear{Harper and Konstan}{Harper and
  Konstan}{2015}]%
        {movielens}
\bibfield{author}{\bibinfo{person}{F.~Maxwell Harper} {and}
  \bibinfo{person}{Joseph~A. Konstan}.} \bibinfo{year}{2015}\natexlab{}.
\newblock \showarticletitle{The MovieLens Datasets: History and Context}.
\newblock \bibinfo{journal}{\emph{ACM Trans. Interact. Intell. Syst.}}
  \bibinfo{volume}{5}, \bibinfo{number}{4}, Article \bibinfo{articleno}{19}
  (\bibinfo{date}{Dec.} \bibinfo{year}{2015}), \bibinfo{numpages}{19}~pages.
\newblock
\showISSN{2160-6455}
\urldef\tempurl%
\url{https://doi.org/10.1145/2827872}
\showDOI{\tempurl}


\bibitem[\protect\citeauthoryear{Knees, Deldjoo, Moghaddam, Adamczak, Leyson,
  and Monreal}{Knees et~al\mbox{.}}{2019}]%
        {recsys-challenge-2019-session-based-hotel-recommendations}
\bibfield{author}{\bibinfo{person}{Peter Knees}, \bibinfo{person}{Yashar
  Deldjoo}, \bibinfo{person}{Farshad~Bakhshandegan Moghaddam},
  \bibinfo{person}{Jens Adamczak}, \bibinfo{person}{Gerard-Paul Leyson}, {and}
  \bibinfo{person}{Philipp Monreal}.} \bibinfo{year}{2019}\natexlab{}.
\newblock \showarticletitle{RecSys Challenge 2019: Session-Based Hotel
  Recommendations}. In \bibinfo{booktitle}{\emph{Proceedings of the 13th ACM
  Conference on Recommender Systems}} (Copenhagen, Denmark)
  \emph{(\bibinfo{series}{RecSys '19})}. \bibinfo{publisher}{Association for
  Computing Machinery}, \bibinfo{address}{New York, NY, USA},
  \bibinfo{pages}{570--571}.
\newblock
\showISBNx{9781450362436}
\urldef\tempurl%
\url{https://doi.org/10.1145/3298689.3346974}
\showDOI{\tempurl}


\bibitem[\protect\citeauthoryear{Lee, Lakshmanan, Tiwari, and Shah}{Lee
  et~al\mbox{.}}{2014}]%
        {modeling-impression-discounting-in-large-scale-recommender-systems}
\bibfield{author}{\bibinfo{person}{Pei Lee}, \bibinfo{person}{Laks~V.S.
  Lakshmanan}, \bibinfo{person}{Mitul Tiwari}, {and} \bibinfo{person}{Sam
  Shah}.} \bibinfo{year}{2014}\natexlab{}.
\newblock \showarticletitle{Modeling Impression Discounting in Large-Scale
  Recommender Systems}. In \bibinfo{booktitle}{\emph{Proceedings of the 20th
  ACM SIGKDD International Conference on Knowledge Discovery and Data Mining}}
  (New York, New York, USA) \emph{(\bibinfo{series}{KDD '14})}.
  \bibinfo{publisher}{Association for Computing Machinery},
  \bibinfo{address}{New York, NY, USA}, \bibinfo{pages}{1837--1846}.
\newblock
\showISBNx{9781450329569}
\urldef\tempurl%
\url{https://doi.org/10.1145/2623330.2623356}
\showDOI{\tempurl}


\bibitem[\protect\citeauthoryear{Paudel, Christoffel, Newell, and
  Bernstein}{Paudel et~al\mbox{.}}{2016}]%
  {updatable-accurate-diverse-and-scalable-recommendations-for-interactive-applications}
\bibfield{author}{\bibinfo{person}{Bibek Paudel}, \bibinfo{person}{Fabian
  Christoffel}, \bibinfo{person}{Chris Newell}, {and} \bibinfo{person}{Abraham
  Bernstein}.} \bibinfo{year}{2016}\natexlab{}.
\newblock \showarticletitle{Updatable, Accurate, Diverse, and Scalable
  Recommendations for Interactive Applications}.
\newblock \bibinfo{journal}{\emph{ACM Trans. Interact. Intell. Syst.}}
  \bibinfo{volume}{7}, \bibinfo{number}{1}, Article \bibinfo{articleno}{1}
  (\bibinfo{date}{Dec.} \bibinfo{year}{2016}), \bibinfo{numpages}{34}~pages.
\newblock
\showISSN{2160-6455}
\urldef\tempurl%
\url{https://doi.org/10.1145/2955101}
\showDOI{\tempurl}


\bibitem[\protect\citeauthoryear{Polato and Aiolli}{Polato and Aiolli}{2016}]%
  {a-preliminary-study-on-a-recommender-system-for-the-job-recommendation-challenge}
\bibfield{author}{\bibinfo{person}{Mirko Polato} {and} \bibinfo{person}{Fabio
  Aiolli}.} \bibinfo{year}{2016}\natexlab{}.
\newblock \showarticletitle{A Preliminary Study on a Recommender System for the
  Job Recommendation Challenge}. In \bibinfo{booktitle}{\emph{Proceedings of
  the Recommender Systems Challenge}} (Boston, Massachusetts, USA)
  \emph{(\bibinfo{series}{RecSys Challenge '16})}.
  \bibinfo{publisher}{Association for Computing Machinery},
  \bibinfo{address}{New York, NY, USA}, Article \bibinfo{articleno}{1},
  \bibinfo{numpages}{4}~pages.
\newblock
\showISBNx{9781450348010}
\urldef\tempurl%
\url{https://doi.org/10.1145/2987538.2987549}
\showDOI{\tempurl}


\bibitem[\protect\citeauthoryear{Rendle, Freudenthaler, Gantner, and
  Schmidt-Thieme}{Rendle et~al\mbox{.}}{2009}]%
        {bpr-bayesian-personalized-ranking-from-implicit-feedback}
\bibfield{author}{\bibinfo{person}{Steffen Rendle}, \bibinfo{person}{Christoph
  Freudenthaler}, \bibinfo{person}{Zeno Gantner}, {and} \bibinfo{person}{Lars
  Schmidt-Thieme}.} \bibinfo{year}{2009}\natexlab{}.
\newblock \showarticletitle{BPR: Bayesian Personalized Ranking from Implicit
  Feedback}. In \bibinfo{booktitle}{\emph{Proceedings of the Twenty-Fifth
  Conference on Uncertainty in Artificial Intelligence}} (Montreal, Quebec,
  Canada) \emph{(\bibinfo{series}{UAI '09})}. \bibinfo{publisher}{AUAI Press},
  \bibinfo{address}{Arlington, Virginia, USA}, \bibinfo{pages}{452--461}.
\newblock
\showISBNx{9780974903958}


\bibitem[\protect\citeauthoryear{S{\o}rensen}{S{\o}rensen}{1948}]%
        {dice-similarity-1}
\bibfield{author}{\bibinfo{person}{T.J. S{\o}rensen}.}
  \bibinfo{year}{1948}\natexlab{}.
\newblock \bibinfo{booktitle}{\emph{A Method of Establishing Groups of Equal
  Amplitude in Plant Sociology Based on Similarity of Species Content and Its
  Application to Analyses of the Vegetation on Danish Commons}}.
\newblock \bibinfo{publisher}{I kommission hos E. Munksgaard}.
\newblock
\showLCCN{a49002195}
\urldef\tempurl%
\url{https://books.google.it/books?id=rpS8GAAACAAJ}
\showURL{%
\tempurl}


\bibitem[\protect\citeauthoryear{Tversky}{Tversky}{1977}]%
        {tversky-features-of-similarity}
\bibfield{author}{\bibinfo{person}{Amos Tversky}.}
  \bibinfo{year}{1977}\natexlab{}.
\newblock \showarticletitle{Features of similarity.}
\newblock \bibinfo{journal}{\emph{Psychological review}} \bibinfo{volume}{84},
  \bibinfo{number}{4} (\bibinfo{year}{1977}), \bibinfo{pages}{327}.
\newblock


\bibitem[\protect\citeauthoryear{Wang and He}{Wang and He}{2018}]%
        {click-through-rate-estimates-based-on-deep-learning}
\bibfield{author}{\bibinfo{person}{Wentao Wang} {and} \bibinfo{person}{Dongzhi
  He}.} \bibinfo{year}{2018}\natexlab{}.
\newblock \showarticletitle{Click-through Rate Estimates Based on Deep
  Learning}. In \bibinfo{booktitle}{\emph{Proceedings of the 2018 2nd
  International Conference on Deep Learning Technologies}} (Chongqing, China)
  \emph{(\bibinfo{series}{ICDLT '18})}. \bibinfo{publisher}{Association for
  Computing Machinery}, \bibinfo{address}{New York, NY, USA},
  \bibinfo{pages}{12--15}.
\newblock
\showISBNx{9781450364737}
\urldef\tempurl%
\url{https://doi.org/10.1145/3234804.3234811}
\showDOI{\tempurl}


\end{thebibliography}

\end{document}